\documentstyle[12pt]{article}
\title{On quantum model of supergravity compensator}
\author{I.L.Buchbinder and A.Yu.Petrov\\
      {\small\it{Department of Theoretical Physics}}\\
      {\small\it{Tomsk State Pedagogical University}}\\
      {\small\it{Tomsk 634041, Russia}}}
\date{}
\begin{document}
\maketitle
\begin{abstract}
A new $N=1$ superfield model in $D=4$ flat superspace is suggested. This
model describes dynamics of chiral compensator and can be treated as a
low-energy limit of $D=4$, $N=1$ quantum superfield supergravity.
Renormalization structure of this model is studied and one-loop counterterms
are calculated. It is shown that the theory is infrared free.
An effective action for the model under consideration is
investigated in infrared domain. The lower contributions to the
one-loop effective action are computed in explicit form.
\end{abstract}

1. It is well known that four-dimensional simple supergravity can be formulated
in $N=1$ curved superspace in terms of vector superfield $H^m(z)$ and chiral
$\Phi(z)$ and antichiral $\bar{\Phi}(z)$ compensators [1-3] (see also [4]). Here
as usual $z^m\equiv(x^m, \theta^{\mu}, \bar{\theta}^{\dot{\mu}})$ are the
superspace coordinates. General approach to quantum superfield supergravity has
been developed in refs. [5,6]. However a number of concrete quantum aspects has
not been investigated in details because of complicated enough structure of the
superfield supergravity itself.

Our goal in the present paper is to formulate a simplified model associated with
$N=1, D=4$ superfield supergravity and describing a dynamics only of compensator
superfields in flat superspace. We show that this model makes possible to carry
out a detailed investigation of many quantum aspects and possessses remarkable
properties in infrared domain. It allows to treat such a model as a natural low-
energy limit of quantum superfield supergravity with matter.

The model under consideration is based on an idea of induced gravity. An approach
to four-dimensional quantum gravity model induced by conformal anomaly of matter
fields in curved space-time has been suggested in ref. [6]. It has been shown
that this model is super-renormalizable in infrared limit and infrared free.
In the given paper we present a supersymmetric model of induced quantum gravity
and investigate its quantum aspects.

2. We start with known superconformal anomaly in $N=1$ curved superspace.
This anomaly is formulated in terms of supercurrent and supertrace [7,8] (see
also [4]) which are the superfield analogs of the energy-momentum tensor and its
trace respectively. For superconformally invariant theories the classical
supertrace $V$ vanishes. However in quantum theory the supertrace $V$ is not
equal to zero because of superconformal anomaly and can be written as follows
[9,10,4]
\begin{equation}
V_A=-{(4\pi)}^{-2}(aW^2 +bG+c(\bar{D}^2 -8R)D^2R)
\end{equation}
Here $a$, $b$, $c$ are some numbers depending on the field content of the theory,
$W^2=W^{\alpha\beta\gamma}W_{\alpha\beta\gamma}$,
$G=W^2-\frac{1}{4}(\bar{D}^2-8R)(G^a G_a+2R\bar{R})$, $D_A=(D_a, D_{\alpha},
\bar{D}_{\dot{\alpha}})$ are the supercovariant derivatives;
$R$, $G_a$, $W_{\alpha\beta\gamma}$ are superfield strengths expressed in terms
of supergravity prepotentials $H^m,\Phi,\bar{\Phi}$;
(we use the denotions accepted in ref. [4]).

A superfield action leading to the anomaly (1) has been found in ref.
[11]. Our goal is to investigate a model a total action of which is a sum
of the anomaly generating action [11] and the action of $N=1, D=4$ superfield
supergravity (see [4]). Being transformed to conformally flat superspace
the total action of our model takes the form

\begin {eqnarray}
S&=&\int d^8z
    (-\frac{Q^2}{2{(4\pi)}^2}\bar{\sigma}\Box\sigma+\bar{D}^{\dot\alpha}\bar
{\sigma}
D^{\alpha}\sigma\times\nonumber\\
&\times&({\xi_1\partial_{\alpha}}_{\dot{\alpha}}(\sigma+\bar{\sigma})
+\xi_2\bar{D}_{\dot{\alpha}}\bar{\sigma} D_{\alpha}\sigma)-
\frac{m^2}{2} e^{\sigma+\bar{\sigma}})+(\Lambda\int d^6z e^{3\sigma}+h.c.)
\end {eqnarray}
Here we use the flat supercovariant derivatives
$D_{\alpha}$, $\bar{D}_{\dot{\alpha}}$, $\partial_{\alpha\dot{\alpha}}$;
$\sigma=\ln\Phi$, $\Lambda$ is a cosmological constant
$m^2=\frac{6}{\kappa^2}$ where $\kappa$ is the gravitational constant;
the $Q^2$, $\xi_1$, $\xi_2$ are expressed in terms of $a$, $b$, $c$.
We will consider the $Q^2$, $\xi_1$, $\xi_2$, $m^2$, $\Lambda$ as the arbitrary
and independent parameters of the model. The superfield model with the
action (2) is the basic object of this paper. We are going to investigate a
renormalization structure of the model (2), compute the one-loop counterterms,
study a behaviour of running couplings and calculate one-loop effective action.

3. To clarify a renormalization structure and calculate the counterterms we
should find the superpropagator and vertices for the model (2).
This model is characterized by the matrix superpropagator with the matrix
elements $G_{++}$, $G_{+-}$, $G_{-+}$, $G_{--}$ where the sign $+$ corresponds
to chiral field and the sign $-$ to the antichiral one. The explicit form of
above matrix is given by the quadratic part of the action (2) and can
be found easy enough. The model under consideration has the vertices of
the four types:  $V_1=\xi_1 \bar{D}^{\dot{\alpha}}\bar{\sigma}
D^{\alpha}\sigma\partial_{\alpha\dot{\alpha}}(\sigma+\bar{\sigma})$,
$V_2=\xi_2\bar{D}_{\dot{\alpha}}\bar{\sigma}\bar{D}^{\dot{\alpha}}\bar{\sigma}
D^{\alpha}\sigma D_{\alpha}\sigma$,
$V_3=-\frac{m^2}{2}(e^{\sigma+\bar{\sigma}}-1-(\sigma+\bar{\sigma})-\frac{1}{2}
{(\sigma+\bar{\sigma})}^2)$,
$V_4=\Lambda (-\frac{D^2}{4\Box})(e^{3\sigma}-1-3\sigma-\frac{9}{2}\sigma^2)+
\bar{\Lambda} (-\frac{\bar{D}^2}{4\Box})(e^{3\bar{\sigma}}-1-3\bar{\sigma}
-\frac{9}{2}\bar{\sigma}^2)$.
All vertices correspond to the action written as an integral over whole
superspace.

A renormalization structure of the theory is defined in terms of superficial
degree of divergences $\omega$.
For explicit calculation of the $\omega$ and the counterterms we will use the
supergraph technique (see [12,4]) specially generalized for use namely in our
model (2).
A value of the $\omega$ is caused by the $D$-factors
and momentum dependence of the superpropagator, by the form of vertices and by
integration over momenta in the supergraphs.
The total contribution to $\omega$ from superpropagators, vertices and
integrations looks like this $\omega\leq 2V_{1,2}-2L_2-5L_1+2L-V_4$
where $V_{1,2,4}$ are numbers of vertices of the type $V_{1,2}$ and $V_4$
respectively in supergraph, $L_1$ is a number $G_{++}$, $G_{--}$-lines,
$L_2$ is a number of $G_{+-}$-, $G_{-+}$-lines and $L$ is a number of
loops in the supergraph.  Taking into account the identities $L+V-P=1$,
$V=V_1+V_2+V_3+V_4$ where $P$ is a number of all internal lines we
obtain the final result for $\omega$.  \begin {equation} \omega\leq
2-3L_1-2V_3-3V_4 \end {equation} The condition of divergence
 $\omega\geq 0$ leads to $L_1=0, V_4=0, V_3=0,1$.  Hence the divergent
diagrams cannot contain the $V_4$-type vertices and $G_{++}$-,
$G_{--}$-type lines. They can include no more that one vertex of
$V_3$-type.
We see a number of divergent structures is more restricted in compare with
non-supersymmetric case [6]. In particular we have a non-renormalization
theorem: the vertex of $V_4$-type is always finite.

Thus, the divergent supergraphs can include an arbitrary
number of $G_{+-}$-, $G_{-+}$-lines and arbitrary number of $V_{1,2}$-type
vertices. But if $\xi_1=\xi_2=0$ then the vertices of $V_{1,2}$-
type are absent at all. The only vertices which can present in divergent
diagrams are now the $V_3$-type ones. But their number is exactly equal to 1
and corresponds only to one-loop supergraphs. It means, at $\xi_1=\xi_2=0$ the
theory becomes to be super-renormalizable. We will show that in infrared limit
the running couplings $\xi_1(t)$ and $\xi_2(t)$ tend to zero and namely above
case is realized.

4. Let us investigate the one-loop renormalizability of the model (2). We start
with calculations of the renormalization constants for the parameters
$\xi_1$, $\xi_2$ and $Q^2$. The corresponding supergraphs are given by the 
Fig.1, Fig.2 and Fig.3

\begin{center}
\begin{picture}(100,100)
\put(50,50){\circle{40}}
\put(30,50){\line(-1,-1){20}}
\put(15,40){\line(0,-1){10}}
\put(10,20){$D^{\alpha}$}
\put(35,70){\line(0,-1){10}}
\put(30,70){$\bar{D}_{\dot{\alpha}}$}
\put(30,50){\line(-1,1){20}}
\put(15,70){\line(0,-1){10}}
\put(35,40){\line(0,-1){10}}
\put(30,20){$D_{\alpha}$}
\put(10,75){$\bar{D}^{\dot{\alpha}}$}
\put(70,50){\line(1,0){20}}
\put(80,40){$\partial_{\beta\dot{\beta}}$}
\put(80,55){\line(0,-1){10}}
\put(65,40){\line(0,-1){10}}
\put(65,70){\line(0,-1){10}}
\put(70,75){$D^{\beta}$}
\put(70,20){$\bar{D}^{\dot{\beta}}$}
\put(45,80){$G_{+-}$}
\put(45,20){$G_{-+}$}
\put(40,0){Fig.1}
\end{picture}
\begin{picture}(100,100)
\put(50,50){\circle{40}}
\put(30,50){\line(-1,-1){20}}
\put(30,50){\line(-1,1){20}}
\put(70,50){\line(1,-1){20}}
\put(70,50){\line(1,1){20}}
\put(15,40){\line(0,-1){10}}
\put(10,20){$D^{\alpha}$}
\put(15,70){\line(0,-1){10}}
\put(10,75){$\bar{D}^{\dot{\alpha}}$}
\put(85,70){\line(0,-1){10}}
\put(85,40){\line(0,-1){10}}
\put(85,20){$D^{\beta}$}
\put(85,75){$\bar{D}^{\dot{\beta}}$}
\put(35,40){\line(0,-1){10}}
\put(30,20){$D_{\alpha}$}
\put(35,70){\line(0,-1){10}}
\put(30,70){$\bar{D}_{\dot{\alpha}}$}
\put(65,40){\line(0,-1){10}}
\put(65,70){\line(0,-1){10}}
\put(70,75){$D_{\beta}$}
\put(70,20){$\bar{D}_{\dot{\beta}}$}
\put(45,80){$G_{+-}$}
\put(45,20){$G_{-+}$}
\put(40,0){Fig.2}
\end{picture}

\begin{picture}(100,100)
\put(50,50){\circle{40}}
\put(30,50){\line(-1,0){20}}
\put(70,50){\line(1,0){20}}
\put(15,55){\line(0,-1){10}}
\put(80,55){\line(0,-1){10}}
\put(80,40){$\partial_{\alpha\dot{\alpha}}$}
\put(10,60){$\partial_{\beta\dot{\beta}}$}
\put(65,40){\line(0,-1){10}}
\put(65,70){\line(0,-1){10}}
\put(70,75){$D^{\alpha}$}
\put(70,20){$\bar{D}^{\dot{\alpha}}$}
\put(35,40){\line(0,-1){10}}
\put(30,20){$D^{\beta}$}
\put(35,70){\line(0,-1){10}}
\put(30,70){$\bar{D}^{\dot{\beta}}$}
\put(45,80){$G_{+-}$}
\put(45,20){$G_{-+}$}
\put(40,0){Fig.3}
\end{picture}
\begin{picture}(100,100)
\put(50,50){\circle{40}}
\put(30,50){\line(-1,-1){20}}
\put(30,50){\line(-1,1){20}}
\put(20,45){$\vdots$}
\put(75,50){$G_{+-}$}
\put(40,0){Fig.4}
\end{picture}
\end{center}

The straightforward calculations of the divergent parts of these supergraphs
within dimensional reduction scheme (see f.e. [4]) allow to find the
renormalization transformations in the form
\begin{eqnarray}
Q^2_{(0)}&=&\mu^{-\epsilon} Z_Q Q^2,\ \ \xi_{(0) 1,2}= \mu^{-\epsilon}
Z \xi_{1,2}\\ Z_Q&=&1+\frac{2^{14} 3^2 \pi^6 \xi_1^2}{Q^6\epsilon},\ \
Z=1+\frac{2^{11} 3^2\pi^2\xi_2}{Q^4\epsilon}\nonumber \end{eqnarray}
where $\epsilon=4-d$ is a parameter of dimensional regularization and $\mu$
is an arbitrary parameter of mass dimension, $Q^2_{(0)}$, $\xi_{1(0)}$,
$\xi_{2(0)}$ are the bare parameters and $Q^2$, $\xi_1$, $\xi_2$ are the
renormalized ones. We see that in one-loop approximation there is the same
renormalization constant $Z$ both for $\xi_1$ and $\xi_2$. It means in
particular, if we put $c\xi_{1(0)}=\xi_{2(0)}$ with some constant $c$ then
the renormalized parameters $\xi_1$ and $\xi_2$ satisfy the same relation
$c\xi_1=\xi_2$. One-loop renormalization does not destroy relationship between
these parameters.

Let us consider a renormalization of the parameter $m^2$ at $\xi_1=\xi_2=0$.
As we will see, namely this situation is realized in infrared limit.
The divergent supergraphs contributing to renormalization of $m^2$
include the only vertex of $V_3$-type, the only $G_{+-}$-type internal
line and an arbitrary number of the external lines of the $\sigma$ and
$\bar{\sigma}$ superfields.  The corresponding supergraphs are given by
Fig.4. The straightforward calculations lead to \begin{equation} m^2_0=
Z_{m^2} m^2;\ \ Z_{m^2}=1+\frac{2}{Q^2\epsilon} \end{equation} Here
$m^2_0$ is a bare parameter and $m^2$ is a renormalized one.

The eqs. (4,5) define the one-loop renormalization of the model. As for the
parameter $\Lambda$ in (2), it was already noted that all supergraphs
containing the vertices of $V_4$-type are finite. It means that the
coupling $\Lambda$ is not renormalized. The fields $\sigma$ and
$\bar{\sigma}$ are not renormalized as well.

5. We consider the behaviour of the running couplings in the model (2).
The corresponding renormalization group equations can be found on the base of
(4,5).

The equations for running constants $\xi_1(t)$, $\xi_2(t)$, $Q^2(t)$ look
like this
\begin{eqnarray}
\frac{d\xi_1}{dt}=a\frac{\xi_1\xi_2}{Q^4};\
 \frac{d\xi_2}{dt}=a\frac{\xi_2^2}{Q^4};\
 \frac{d Q^2}{dt}=b\frac{\xi_1^2}{Q^4}
\end{eqnarray}
where $a=2^{11}3^2\pi^2$, $b=3^2 2^{14}\pi^4$.
The solutions of these equations have the form
\begin{eqnarray}
\xi_1(t)&=&\frac{\xi_1}{\xi_2}\xi_2(t)\nonumber\\
Q^2(t)&=& Q^2+8\pi^2\frac{\xi_1^2}{\xi_2^2}(\xi_2(t)-\xi_2)\nonumber\\
t&=&\frac{1}{2^{11} 3^2\pi^2}\Big\{-[Q^2-8\pi^2\frac{\xi^2_1}{\xi^2_2}]
(\frac{1}{\xi_2(t)}-\frac{1}{\xi_2})-\\
&-&16\pi^2{(\frac{\xi_1}{\xi_2})}^2 [Q^2-8\pi^2\frac{\xi^2_1}{\xi^2_2}]
\ln\frac{\xi_2(t)}{\xi_2}+
64\pi^4{(\frac{\xi_1}{\xi_2})}^4(\xi_2(t)-\xi_2)\Big\}\nonumber
\end{eqnarray}
Here $\xi_{1,2}=\xi_{1,2}(t)|_{t=0}$, $Q^2=Q^2(t)|_{t=0}$. It is easy to see
that in infrared limit $t\rightarrow-\infty$ the $\xi_1^{(0)}= \xi_2^{(0)}=0$
is an infrared fixed point and $Q^2(t)\to Q^2$. This result shows that
only supergraphs given by Fig.4 can contribute to $m^2$-renormalization
in infrared limit.

Now we consider the running coupling for the parameter $m^2$ in infrared domain
$\xi_1=\xi_2=0$. Following to ref.[6] we make the scale transformation
$\sigma\rightarrow\alpha\sigma$,
$\bar{\sigma}\rightarrow\alpha\bar{\sigma}$ and
$S\rightarrow\frac{1}{\alpha^2}S$
where $\alpha$ is a real parameter. The only modification in compare with above
consideration is that we should transform the propagator
$G_{+-}\to \alpha^2G_{+-}$. It leads to
\begin{eqnarray}
m^2_{(0)}=Z_{m^2} m^2,\ \
Z_{m^2}=1+\frac{2\alpha^2}{Q^2\epsilon}\nonumber \end{eqnarray} This
relation leads to the following equation for running $m^2(t)$
\begin{equation}
\frac{d m^2(t)}{dt}=\frac{2\alpha^2}{Q^2}m^2(t)+\Delta_{m^2} m^2(t)
\end{equation}
where $\Delta_{m^2}$ is a scaling dimension of $m^2$. To find $\Delta_{m^2}$
we consider the term $\frac{m^2}{2\alpha^2} e^{\alpha(\sigma+\bar{\sigma})}$
in the action (2) after above scale transformation of the fields $\sigma$,
$\bar{\sigma}$ and action $S$. Scaling dimension of this term is $-2$ (we take
into account that the action is dimensionless), scaling dimension of
$e^{\alpha(\sigma+\bar{\sigma})}$ is $2\alpha$. Hence $\Delta_{m^2}=2-2\alpha$.
Therefore, the eq. (8) looks like this
\begin{eqnarray}
\frac{d m^2(t)}{dt}&=&(2-2\alpha+\frac{2\alpha^2}{Q^2})m^2(t)
\end{eqnarray}
Hence
\begin{equation}
m^2(t)=m^2\exp((2-2\alpha+\frac{2\alpha^2}{Q^2})t)
\end{equation}
where $m^2=m^2(t)|_{t=0}$. It is evident that at
$2-2\alpha+\frac{2\alpha^2}{Q^2}>0$
we get $m^2(t)\rightarrow 0$ i.e the running parameter $m^2(t)$ vanishes in
infrared limit.

Taking into account everything above we can say that the model with the action
\begin{eqnarray}
S=\int d^8 z(-\frac{1}{2}\frac{Q^2}{16\pi^2}\bar{\sigma}\Box\sigma)
+(\frac{\Lambda}{\alpha^2}\int d^6 z e^{3\alpha\sigma}+ h.c.)
\end{eqnarray}
is a natural infrared limit of the model (2). The model (11) corresponds
to finite quantum theory, i.e. all quantum corrections to effective action in
this model will be finite.

6. Let us consider the effective action corresponding to the model (2)
in infrared limit where  $\xi_1=\xi_2=m^2=0$.
As it has been proved in previous sections the model (2) will be finite in
infrared domain and we face a problem of effective action in finite quantum
field theory.

Aspects of effective action in supersymmetric models have been investigated in
a number of papers (see f.e. [13-19]). 
\footnote {We would like to notice that some classical aspects of the 
first paper in ref.[13] have also been considered in earlier paper [30].}
We will follow our approach [18, 19]
based on superfield generalization of proper-time method.

We suggest that the effective action has the following general structure
\begin{equation}
\Gamma=\int d^8 z L+(\int d^6 z {\cal L}_c +h.c.)
\end{equation}
Here $L$ can be called the general effective lagrangian,
${\cal L}_c$ can be called the chiral effective lagrangian.
The $L$ should depend on the field $\sigma$, $\bar{\sigma}$ and their
supercovariant derivatives of any order.
In particular there can be a term independent of the derivatives. We will call
this term the kahlerian effective potential $K$ (see the discussion in [18]).
The ${\cal L}_c$ is a chiral superfield and therefore it can depend only on
$\sigma$ and its space-time derivatives of any order.

To calculate the effective action we will use a technique of loop expansion.
In particular, in order to find one-loop contribution to the effective action we
should divide the fields into background and quantum and consider a quadratic
in quantum fields part of the action depending on background fields (see the
details f.e. in [20]).

The effective lagrangians $L(\sigma,\bar{\sigma})$ and ${\cal L}_c(\sigma)$ can
be represented in the form of power expansion in derivatives of $\sigma$,
$\bar{\sigma}$. The lower correction to $L$ should has the form
$L \sim \Lambda^{k_1} \bar{\Lambda}^{k_2} e^{n(\sigma+\bar{\sigma})}$ where
$k_1, k_2$ and $n$ are some numbers. It is turned out that all these numbers
can be found exactly without any calculations.

We notice that the action (2) at $\xi_1=\xi_2=m^2=0$ is
invariant under transformations
$\sigma\to \sigma+\gamma$, $\bar{\sigma}\to \bar{\sigma}+\beta$,
$\Lambda\to e^{-3\gamma}\Lambda$, $\bar{\Lambda}\to e^{-3\beta}\Lambda$
where $\gamma$ and $\beta$ are constant independent parameters.
The effective action should be also invariant under these transformations
(the theory under consideration is finite quantum theory therefore we have no 
any
sources of anomalies there). It means $-3k_1+n=0$, $k_2=k_1$. Since the
dimension of $L$ is 2 we get $3(k_1+k_2)=2$. As a result we find $k_1=k_2=1/3$,
$n=1$. Hence the lower contribution to kahlerian effective potential looks like
\begin{equation}
K= \rho {(\lambda\bar{\lambda})}^{1/3} e^{\sigma+\bar{\sigma}}
\end{equation}
where $\rho$ is some dimensionless constant which
can be found only at straightforward calculations.

The lower contribution to ${\cal L}_c$ should has the form of a linear
combination of the terms\\
$\Lambda^{q_1}\bar{\Lambda}^{q_2}e^{p\sigma}
\partial_m\sigma\partial^m\sigma$ and
$\Lambda^{q_1}\bar{\Lambda}^{q_2}e^{p\sigma}\Box\sigma$
where $q_1$, $q_2$, $p$ are some numbers. The dimension of ${\cal L}_c$
is 3, hence $3(q_1+q_2)+2=3$. The invariance of ${\cal L}_c$ under above
transformations leads to $-3q_1+p=0$ (we use transformations at $\beta=0$
there). As a result $q_1=p/3$, $q_2=1/3-p/3$. Hence the lower contribution
to ${\cal L}_c$ looks like this
\begin{equation}
{\cal L}_c=\bar{\Lambda}^{1/3}\big[
\sum_{p_1} {(\frac{\Lambda}{\bar{\Lambda}})}^{p_1} \rho_{p_1} e^{p_1\sigma}
\partial_m\sigma\partial^m\sigma+ \sum_{p_2} {(\frac{\Lambda}
{\bar{\Lambda}})}^{p_2} \rho_{p_2} e^{p_2\sigma}\Box\sigma\big]
\end{equation}
where $\rho_{p_1}$ and $\rho_{p_2}$ are some dimensionless constants. These
constants and the values of $p_1$ and $p_2$ can be found at straightforward
calculations.

The one-loop contrbution $\bar{\Gamma}^{(1)}$ to effective action
is given by the relation
$$\exp(i\Gamma^{(1)})=\int D\chi D\bar{\chi}
\exp(iS^{(2)}[\sigma, \bar{\sigma};\chi, \bar{\chi}])$$
$$S^{(2)}=\int d^8 z \big( -\frac{Q^2}{2{(4\pi)}^2}\bar{\chi}\Box\chi\big)+
\big(\frac{9\Lambda}{2}\int d^6 z e^{3\sigma}\chi^2+ h.c.\big)$$
Using the trick suggested in ref. [18] one can rewrite the $\bar{\Gamma}^{(1)}$
in the form
\begin{eqnarray}
\bar{\Gamma}^{(1)}=-\frac{i}{2}\int_0^{\infty} \frac{d\tau}{\tau} sTr
e^{i\tau\Delta}\\ \Delta=-\frac{Q^2}{{(4\pi)}^2}\Box^2-9\Lambda
e^{3\sigma}\frac{\bar{D}^2}{4}- 9\bar{\Lambda}
e^{3\bar{\sigma}}\frac{D^2}{4}\nonumber \end{eqnarray} where $sTr$ is a
functional supertrace of the operator acting on superfields.  We should
like to notice especially that the integral over proper time in (15)
does not need in regularization since the theory under consideration is finite
quantum field theory.

To calculate the one-loop kahlerian effective potential it is sufficient to
consider fields $\sigma$ and $\bar{\sigma}$ as independent of space-time
coordinates. It leads to
$$\exp (i\tau\Delta)=
\exp[i\tau(-\frac{9}{4}\Lambda e^{3\sigma}\bar{D}^2-
\frac{9}{4}\bar{\Lambda}e^{3\bar{\sigma}}D^2)]
\exp(-i\tau\frac{Q^2}{{(4\pi)}^2}\Box^2)$$
In this approximation the functional supertrace in (15) can be exactly
calculated. The final result for kahlerian effective potential has the form
\begin{equation}
K^{(1)}= c{\Big(\frac{\Lambda\bar{\Lambda}}{Q^4}\Big)}^{1/3}
e^{\sigma+\bar{\sigma}}
\end{equation}
where $c$ is a positive constant$^2$. We see the $K^{(1)}$ (16)
completely corresponds to (13) with explicit value for $\rho$.

The calculation of chiral contribution to effective action is more
complicated. To do that we can put $\bar{\sigma}=0$ in operator $\Delta$ (15)
but we should take into account whole superspace dependence of the chiral
superfield $\sigma$. Nevertheless the lower contribution to one-loop chiral
effective lagrangian can be calculated in explicit form
\begin{eqnarray}
{\cal L}^{(1)}_c&=& \frac{\Lambda^{1/3}}{Q^{2/3}}
\{( c_1 e^{-\sigma} +c_2 e^{2\sigma} +c_3 e^{-\sigma} +c_4 e^{-4\sigma})\times\\
&\times&\partial^m\sigma \partial_m\sigma+
\Box\sigma (c_3 e^{-\sigma} +c_4 e^{-4\sigma})\}\nonumber
\end{eqnarray}
where $c_1$, $c_2$, $c_3$, $c_4$ are some definite numbers
\footnote{The explicit forms for the numbers $c$, $c_1$, $c_2$, $c_3$, $c_4$ (16,
17) are given in terms of some power series and look like too cumbersome to be
given there. The details of calculations leading to (16, 17) will be presented
somewhere.}.
 We put for simplicity
$\bar{\Lambda}=\Lambda$ there and further. It is evident that the
${\cal L}^{(1)}_c$ (17) completely corresponds to (14) with explicit values for
the $\rho_{p_1}$, $\rho_{p_2}$.

We want call attention that both $K$ (13) and ${\cal L}_c$ (14) display non-
polynomial dependence on the couplings $\Lambda$ and $\bar{\Lambda}$.

7. As a result we can write the one-loop effective action $\Gamma^{(1)}$ with
lower contributions to $L$ and ${\cal L}_c$. The $\Gamma^{(1)}$ is a sum of
classical action (2) at $\xi_1$= $\xi_2$= $m^2=0$ and the quantum corrections
$\int d^8 z K^{(1)}$ (16) and $(\int d^6 z {\cal L}_c^{(1)}+ h.c.)$ (17).
Let us investigate a structure of this $\Gamma^{(1)}$
for the superfields $\sigma$, $\bar{\sigma}$ slowly varying in
space-time.  It means we can put $\partial_m\sigma\approx 0$,
$\partial_m\bar{\sigma}\approx 0$. If we transform
$\sqrt{c}{\big(\frac{\Lambda}{Q^2}\big)}^{1/3} e^{\sigma}=\phi$,
$\sqrt{c}{\big(\frac{\Lambda}{Q^2}\big)}^{1/3} e^{\bar{\sigma}}=\bar{\phi}$,
we obtain
\begin{equation}
\Gamma=\int d^8 z \phi\bar{\phi}+\big(\lambda \int d^6 z {\phi}^3+ h.c.)
\end{equation}
where $\lambda=Q^2 c^{-3/2}$.
This action describes a dynamics of a massless chiral superfield $\phi$ of
dimension equal to 1 and coincides with the standard action of Wess-
Zumino model. So, we conclde that the model of supergravity chiral compensator
leads for slowly varying superfields to standard Wess-Zumino model.
One can see that the parameter $\Lambda$ is absent in (18), its role
is to transform the dimensionless field $e^{\sigma}$ into the field $\phi$
with dimension 1.

8.  Let us summarize our results.
We have formulated a new model of chiral superfield in $N=1$, $D=4$ flat
superspace. This model is generated by superconformal anomaly of matter
superfields and can be considered as a simplified model of quantum supergravity
in low-energy domain. The features of the model are its complete superfield
formulation, non-trivial interactions of chiral and antichiral superfields,
presence of five couplings, three of which are dimensionless, and higher
(second) derivatives in a kinetic term. The model is a natural
supersymmetric generalization of the low-energy quantum gravity model
given by Antoniadis and Mottola [6].

The analysis of superficial degree of divergence shows that this model leads to
decrease of number of possible divergent structures in comparison with non-
supersymmetric model [6]. We have calculated the one-loop superfield
counterterms, investigated the equations for running couplings and showed that
the model is infrared free and moreover it is finite in infrared limit.

An interesting feature of the model is a non-renormalization theorem according
to which the vertex $\Lambda\int d^6 z e^{3\sigma}$ has no divergent corrections.
Since the superfield $\sigma$ is not renormalized in this model we get that the
parameter $\Lambda$ (cosmological constant) is always finite. The analogous
vertex including cosmological constant in non-supersymmetric model [6]
gets divergent corrections. As a result, unlike non-supersymmetric case, the
beta-function for $\Lambda$ is equal to zero in our model. It means the
mechanism leading to vanishing of effective cosmological constant suggested in
ref. [6] will work only if a supersymmetry is violated.

We have investigated a structure of effective action in the model under
consideration in the infrared limit where the model becomes to be finite.
We have shown that the effective action is defined by general effective
lagrangian and chiral effective lagrangian and found a generic form of lower
contributions to these objects within expansions in derivatives.
Generalized proper-time superfield technique allowing to calculate the one-
loop corrections to effective action has been developed and lower contributions
to effective lagrangians have been calculated in explicit form.
We have picked out that the effective action found there is reduced to
standard Wess-Zumino model for slowly varying in space-time superfields.

The various aspects of quantum gravity induced by conformal anomaly of the
matter fields have recently been investigated in two- and four-dimensional
space-time (see f.e. [6, 21-29]). We hope that the suggested model allows to
consider the analogous and new aspects in quantum supergravity.
\vspace{4mm}

{\Large\bf{Acknowledgements}}

The authors are grateful to S.M.Kuzenko and S.D.Odintsov for interesting
discussions on some aspects of the work. 
We are also very thankful to S.J.Gates paying our attention to paper [30]
I.L.B. thanks D.Luest, D.Ebert and
H.Dorn for their hospitality during his visit to Institute of Physics, Humboldt
Berlin University where most part of the work has been fulfilled.
This visit was supported by Deutsche Forschungsgemeinschaft under
contract DFG$-$436 RUS 113.
 The work was supported in part by ISF  and Russian government
under the grant RI1300 and by Russian Foundation for Basic Research under the
project No.94-02-03234.

\end{document}